\documentclass[prl,twocolumn,superscriptaddress,floatfix,letterpaper]{revtex4}
\usepackage{graphicx,amssymb,bm,amsfonts,amsmath,graphics,color}
\newcommand {\degree}{$^{\circ}$}

\begin{document}

\title{Universal {\em waterfall-like} feature in the spectral function of high temperature superconductors}

\author{J. Graf} \affiliation{Materials Sciences Division, Lawrence Berkeley National Laboratory, Berkeley, CA 94720, USA}
\author {G.-H. Gweon$^{**}$}
\affiliation {Department of Physics, University of California Berkeley, CA 94720, USA}
\author{A. Lanzara} \email{alanzara@lbl.gov}
\affiliation{Materials Sciences Division, Lawrence Berkeley National Laboratory, Berkeley, CA 94720, USA} \affiliation{Department of Physics, University of California Berkeley, CA 94720, USA}

\date {\today}

\begin{abstract}
By performing high resolution angle resolved photoemission spectroscopy (ARPES) experiments on four different families of p-type cuprates, over an energy range much bigger than investigated before, we report the discovery of a universal high energy anomaly in the spectral function. This anomaly is characterized by the presence of two new high energy scales $E_1= 0.35-0.45$eV and $E_2=0.8-0.9$eV and the pinning of the main ARPES spectral function along the boundary of a diamond in the momentum space.  This anomaly unveils a missing link between the doped oxygen holes and the quasiparticles, providing a full range of relevant interactions to the high T$_c$ problem.  
\end{abstract}



\maketitle

\paragraph*{}

Uncovering the key interactions, and hence the relevant energy scales that give rise to the dressing of a doped oxygen hole at high energy into a quasiparticle, is one of the main unsolved issues in the high temperature superconductivity field.
In the case of conventional superconductors for example, the identification of the phonon energy scale was a keystone in the understanding of the superconducting mechanism \cite{Bardeen57,MacMillan}.
For high temperature superconductors (HTSCs), although an equivalent low energy scale at 0.04-0.08eV has been already identified (``kink'') \cite{Bogdanov,Lanzara,Johnson,Kaminski,Gromko,Cuk,Gweon}, whether other important interactions contribute to the quasiparticle formation process still remains an open question.
Indeed, the fact that the ARPES kink of p-type cuprates is difficult to understand due to its unusual high energy behavior \cite {Zhou03,Gweon}, in contrast to the kink of other simpler materials  \cite{phononkinkBe,phononkinkMo}, strongly motivates such a question.

\begin{table}
\begin{center}
\begin{tabular}{|c|c|c|c|c|c|}
\hline
sample & T$_c$ (\degree K)   & $h\nu$ (eV)   & polarization \\
\hline \hline
Bi2212 UD  & 64 & 52         & p$_b$\\
\hline
Bi2212 OPT & 91 & 52         & p$_b$\\
\hline
Bi2212 OD  & 65 & 33         & p$_a$\\
\hline
Bi2201 OPT & 32 & 33         & p$_a$\\
\hline
Eu-LSCO OPT & 15 & 55        & p$_b$\\
\hline
Pb-Bi2212 OD & 65 & 60 \& 130        & p$_b$\\
\hline
\end{tabular}
\vspace{1pc}
\caption{Summary of the samples characteristic (family, doping and critical temperature, $T_c$), and of the experimental conditions (photon energy ($h\nu$) and polarization of the light with respect to the sample (see Fig. 1)) used for the data here presented.}
\end{center}
\end{table}

\begin{figure}
\includegraphics[width=19pc]{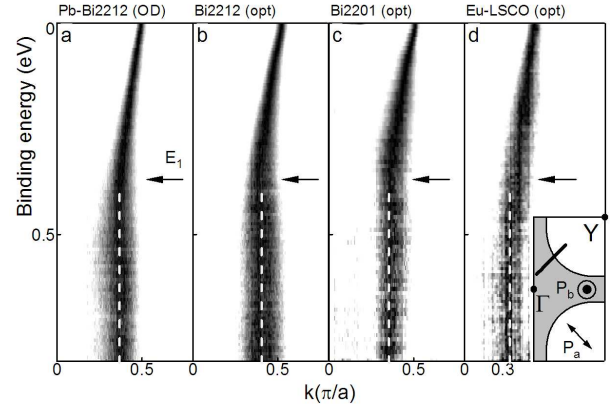}
\label{fig1}
\caption {MDC-normalized maps for four different families of cuprate superconductors at 25K: (a) OD Pb-Bi2212, (b) opt-doped Bi2212, (c) opt-doped Bi2201 and opt- doped Eu-LSCO.  Data are taken along the nodal direction, as shown in the inset of panel d. Black represents maximum of intensity and white zero intensity.  The light polarization was always along the $\Gamma$-X direction (P$_a$), but for some of the data (see table 1), the main component was out of plane (P$_b$)}
\end{figure}

\begin{figure*}
\includegraphics[width=30pc]{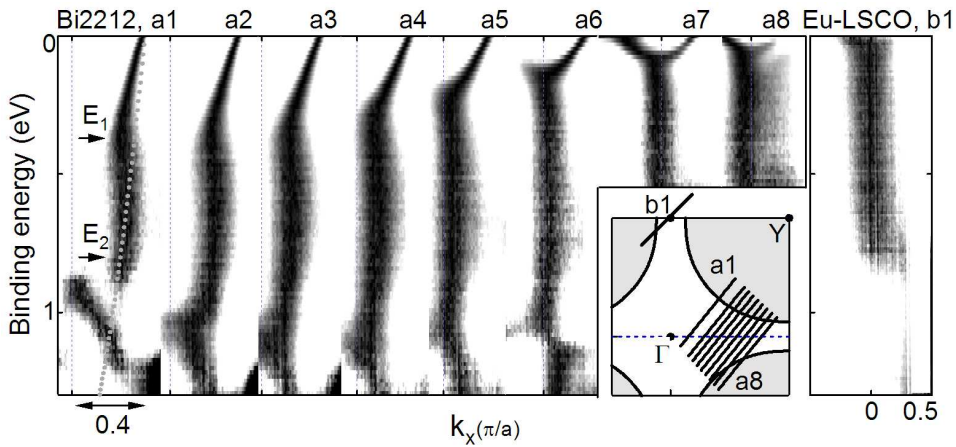}
\label{fig2}
\caption{(a)-(b) MDC-maps for OD-Bi2212 and opt-doped Eu-LSCO.  The momentum position of the cuts $a_1$ to $a_8$ and $b_1$ are shown in the inset. 
The LDA band is shown with a gray dotted line for comparison in panel (a) \cite{LDA}.}
\end{figure*}

In this paper we report the discovery of two universal energy scales, E$_1$= 0.35-0.45eV and E$_2$ =0.8-0.9eV, in the spectral function of p-type cuprates superconductors, identified as the `high energy anomaly' \cite{Jeff1,Jeff2}. E$_1$, is marked by a steep downturn of the main dispersion toward higher energies, like a waterfall of spectral intensity.  E$_2$ is marked by the reappearance of a band-like excitation, dispersing toward $\Gamma$. We provide a simple physical picture behind this anomaly, suggesting that this high energy anomaly bears information on the true nature of the building block of low-energy quasiparticles in cuprates superconductors. The low energy quasiparticle dispersing between E$_F$ and E$_1$ is widely believed to represents the motion of a Zhang Rice Singlet (ZRS)\cite{Zhang} in the antiferromagnetic environment (e.g ZRS coupled to a copper spin). We propose that at E$_1$, the singlet decouples from the copper spin, decaying in a continuum of incoherent excitations, resulting in the observed waterfalls. This continuum of excitations represent the motion of the ZRS.  At E$_2$, the shake off process is complete and only the doped oxygen hole remains, explaining the LDA like dispersion predicted long ago \cite{LDA}. 

High resolution angle resolved photoemission spectroscopy experiments (ARPES) were performed at beamlines 7.0.1, 10.0.1 and 12.0.1 of the Advanced Light Source in Berkeley on single crystals of single layer Bi$_2$Sr$_{1.6}$La$_{0.4}$Cu$_2$O$_{6+x}$ (Bi2201), double layer Bi$_2$Sr$_2$CaCu$_2$O$_{8+x}$ (Bi2212), Pb-doped Bi2212 and optimally doped La$_{1.64}$Eu$_{0.2}$Sr$_{0.16}$CuO$_4$ (Eu-LSCO) over an energy range much bigger than previously investigated \cite{Damascelli,Campuzano}.  The doping of each sample was cross-checked with the nominal value using the measured superconducting-gap value and the binding energy of the Van Hove singularity. Data were taken at different photon energies, polarizations and temperature. The experimental conditions reported in this paper are summarized in Table 1. The total energy and angular resolutions were less than 50meV and less than 0.07$\AA^{-1}$, sufficient for the structures studied here. Except for the Bi2201 data, measured in the first Brillouin zone (BZ), all the data were measured in both the first and the second BZs.
In this paper `low energy' indicates the energy range between $E_F$ and $E_1$ and `high energy' the energy range between $E_1$ and the valence band complex at about 1eV\@.

Figure 1 shows the ARPES intensity vs energy and momentum along the $\Gamma$ to $(\pi,\pi)$ direction, nodal direction for four different p-type cuprates. The background was subtracted and the intensity was normalized to the maximum intensity of the momentum distribution curves (MDC, momentum cuts at fixed energy) for each energy step.  This MDC normalization scheme allows to easily follow the MDC peak position and width as a function of energy without any curve fitting.  In addition to the low energy kink, not apparent within the energy-momentum window used in the figure, a surprising sudden downturn of the dispersion, toward a nearly vertical feature (see white dotted line), occurs at $\approx 0.35$eV  (see arrow). This identifies a new energy scale, $E_1=0.35-0.45$eV, which is also characterized by a sudden decrease of the spectral intensity.  
From now on we will simply refer to this feature as the `waterfall'.  This waterfall extends from E$_1$ to E$_2$ and is characterized by a well defined and almost energy independent peak in the MDCs

The waterfall discussed in Figure 1 persists also in other portion of the BZ, as shown in Figure 2, where we report the MDC-normalized maps for the OD-Bi2212 from the nodal to the near anti-nodal region (cuts $a_0$ to $a_8$) and for the optimally doped Eu-LSCO at the antinodal point, $b_1$.  The MDC-normalization scheme reveals a waterfall-like feature for each cuts. The waterfalls start from the bottom of the low energy dispersion or E$_1$, whichever occurs first along the momentum cut. It is remarkable that the waterfalls are so well localized in the momentum space over a large energy window for each cut.

The LDA band (dotted gray line in panel $a_1$) \cite{LDA}, plotted in the same figure for comparison, shows a good agreement with the experimental data along the nodal direction at an energy of $\approx 0.8-0.9$eV\@.  This allows us to identify a new energy scale $E_2= 0.8-0.9$eV where the MDC peak starts dispersing again toward the $\Gamma$ point. However, we note that it is hard to follow the MDC dispersion all the way to the $\Gamma$ point due to the onset of another dispersing band of high spectral weight, associated with the valence band complex of Bi2212 \cite{Seppoprivate}, whose maximum is at 0.9eV at the $\Gamma$ point.  

In Figure 3 we show the MDC dispersions (solid line), extracted by fitting the MDC curves with Lorentzian, for the OD-Bi2212 (panel a) and Eu-LSCO (panel b) from the nodal to the antinodal direction.  The location of each cut is indicated in the inset of the same figure. 
In the case of Bi2212 we also show the EDC dispersion, extracted from the position of the EDC peak maximum.  An overall good agreement is seen between OD-Bi2212 and Eu-LSCO. However, we observe two main differences. First, the waterfalls in the nodal region of Eu-LSCO are not as steep as for Bi2212. Second, we could not distinguish a peak anymore in the MDCs of the LSCO data at E$_2$. This could however just be due to the lower sample surface quality.

\begin{figure}
\includegraphics[width=20pc]{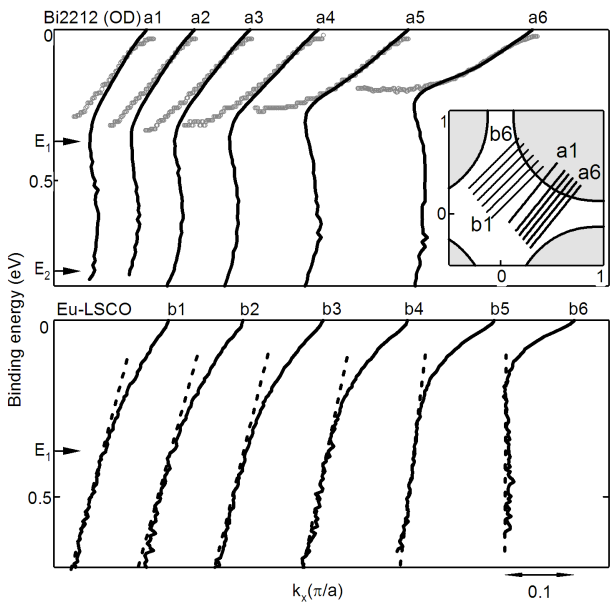}
\label{fig3}
\caption{MDC and EDCs dispersions for the OD-Bi2212 (upper panel) and opt-doped Eu-LSCO (lower panel) from nodal to antinodal direction.  The location of each cuts is indicated in the inset of the upper panel.  Solid line are dispersions extracted from the MDCs peak positions, while grey circles are dispersions extracted from EDC peak positions (upper panel only). The two energy scale $E_1$ and $E_2$ are shown by arrows.}
\end{figure}

We show in figure 4 the full momentum dependence of the waterfalls. The ARPES intensity integrated between $E_F$ and $E_2$ for the four different families of cuprates reported here is shown.  The data reveal the presence of a large momentum region around $\Gamma$ of very low spectral weight, consistent with the sudden decrease of intensity of the EDC peak observed at E$_1$ \cite{Jeff1,Jeff2}.
Note that the spectral weight within this region is {\it not} zero, as shown by the intensity profile along ($-\pi,0$)-($0,0$) (black line).

It is important to point out that the high energy anomaly here discussed cannot be explained in terms of ARPES matrix element, as it is a robust feature of the data independent of photon energy, polarization setting and BZ location.  We note that a similar suppression of the ARPES spectral intensity near the $\Gamma$ point has been reported in the literature for another p-type cuprate \cite{Ronning98}.  

\begin{figure}
\begin{center}
\includegraphics[width=20pc]{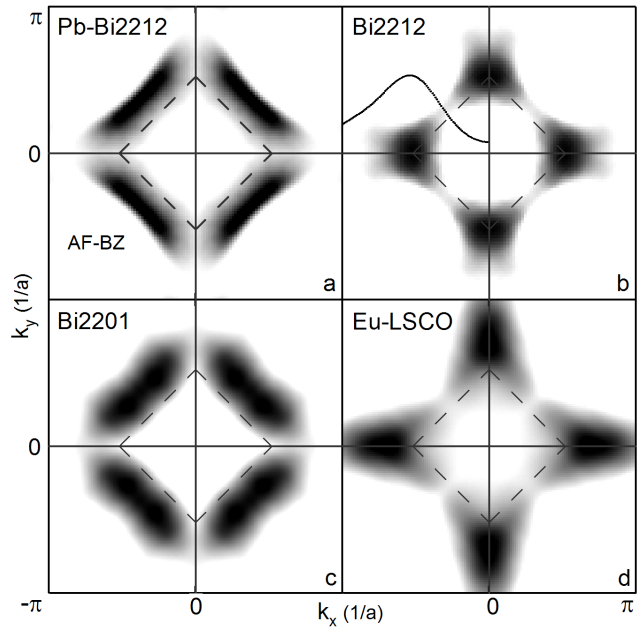}
\label{fig4}
\caption{Panels (a-d) show the ARPES intensity integrated from $E_F$ to 0.8eV\@ for Pb-Bi2212 in the second BZ (panel a), OD-Bi2212 in the first BZ (panel b), opt-doped Bi2201 (panel c) and opt-doped Eu-LSCO in the second BZ zone (panel d). An intensity profile along ($-\pi,0$)-($0,0$) is shown in black in panel b, after subtraction of the elastic background.}
\end{center}
\end{figure}

A possible candidate for the novel behavior presented in this paper is the coupling of electrons to high energy bosons like plasmons.  While in plane plasmons can be a valuable candidate given their high energy \cite{Markievicz}, it is hard to explain the persistence of the high energy anomaly in the undoped compound \cite{Ronning}.  

On the other hand, the material independence of the waterfall and the presence of a similar anomaly in other Mott insulators \cite{Denlinger}, suggest that the two energy scales $E_1$ and $E_2$ represent a complex interplay of the bare oxygen hole with the spin-lattice background.  In addition, the doping independence means that the high energy interactions are at the scale of few lattice constants, since the antiferromagnetic correlation length decreases to two lattice constants when the doping increases to optimal doping and beyond \cite {Kastner98}.

As proposed by Graf et al. \cite{Jeff2}, we believe that the data suggest that the low energy quasiparticle, between $E_F$ and $E_1$ is a local composite fermion made of a  charge $+e$ Zhang Rice singlet (ZRS) and a copper spin $1/2$, bound together to make a composite fermion with the right quantum number to be observed by ARPES. This composite fermions is further dressed by the lattice through a strong electron-phonon interaction, that affects the dispersion from $E_F$ to at least E$_0$ \cite{Gweon,Muller05,Kochelaev97,Bianconi94,Billinge93,Calvani96,Mishchenko04,Shen04}. At $E_1$, the composite fermion breaks down and disintegrates into a ZRS and a copper spin. In the energy range from $E_1$ to $E_2$, the electronic excitations have very little overlap with the photo-hole generated by the photoemission process (as the ZRS does not have the right quantum number to be observed by ARPES). The intensity stems from the tails of the EDCs of the closest band-like excitations.  At energy $E_2$ the ZRS finally disintegrates into a bare oxygen hole and a copper spin, the first of which explains the re-emergence of a band-like dispersion. 

This result suggests the important role of the high energy physics \cite{Phillips,Emery,Kane,Macridin} to determine the true nature of the low energy excitations in the cuprates and together with other universal properties in these materials \cite{Lanzara,Zhou03,Xu00,Bourges00} add another important piece of evidence to the high T$_c$ puzzle. 

$^{**}$ Present address: Department of Physics, University of California, Santa Cruz, CA 95064.

\begin{acknowledgments}
We thank D. H. Lee, A. Bill for useful discussions, and S. I. Uchida, H. Eisaki, H. Takagi and T. Sasagawa for providing us with high quality single crystals for this study. We also thank A. Bostwick and A.V. Fedorov for experimental help.  This work was supported by the National Science Foundation through Grant No. DMR03-49361 and the Director, Office of Science, Office of Basic Energy Sciences, Division of Materials Sciences and Engineering of the U.S Department of Energy under Contract No. DEAC03-76SF00098.  
\end{acknowledgments}

\end{document}